\documentclass[aps,prl,floatfix,superscriptaddress,showpacs,twocolumn]{revtex4}
\usepackage{amsmath}
\usepackage{graphicx}
\usepackage[letterpaper,margin=1in]{geometry}
\newcommand{\prn}[1]{\left(#1\right)}

\def\abrk#1{{\langle#1\rangle}}
\newcommand{\mb}{\mathbf}
\newcommand{\uv}[1]{\mb{\hat{#1}}}
\newcommand{\abs}[1]{\left|#1\right|}

\begin{document}
\title{Can a quantum nondemolition measurement improve the sensitivity of an atomic magnetometer?}
\author{M. Auzinsh}
\affiliation{Department of Physics, University of Latvia, 19
Rainis blvd, Riga, LV-1586, Latvia}
\author{D. Budker}
\email{budker@socrates.berkeley.edu} \affiliation{Department of
Physics, University of California, Berkeley, CA 94720-7300}
\affiliation{Nuclear Science Division, Lawrence Berkeley National
Laboratory, Berkeley CA 94720}
\author{D. F. Kimball}
\affiliation{Department of Physics, University of California,
Berkeley, CA 94720-7300}
\author{S. M. Rochester}
\affiliation{Department of Physics, University of California,
Berkeley, CA 94720-7300}
\author{J. E. Stalnaker}
\affiliation{Department of Physics, University of California,
Berkeley, CA 94720-7300}
\author{A. O. Sushkov}
\affiliation{Department of Physics, University of California,
Berkeley, CA 94720-7300}
\author{V. V. Yashchuk}
\affiliation{Advanced Light Source Division, Lawrence Berkeley
National Laboratory, Berkeley CA 94720}

\date{\today}
\begin{abstract}
Noise properties of an idealized atomic magnetometer that utilizes
spin squeezing induced by a continuous quantum nondemolition
measurement are considered. Such a magnetometer measures spin
precession of $N$ atomic spins by detecting optical rotation of
far-detuned light. Fundamental noise sources include the quantum
projection noise and the photon shot-noise. For measurement times
much shorter than the spin-relaxation time observed in the absence
of light ($\tau_{\rm rel}$) divided by $\sqrt{N}$, the optimal
sensitivity of the magnetometer scales as $N^{-3/4}$, so an
advantage over the usual sensitivity scaling as $N^{-1/2}$ can be
achieved. However, at longer measurement times, the optimized
sensitivity scales as $N^{-1/2}$, as for a usual shot-noise
limited magnetometer. If strongly squeezed probe light is used,
the Heisenberg uncertainty limit may, in principle, be reached for
very short measurement times. However, if the measurement time
exceeds $\tau_{\rm rel}/N$, the $N^{-1/2}$ scaling is again
restored.
\end{abstract}
\pacs{33.55.Ad,42.50.Lc,07.55.Ge}

%32.60.+i Zeeman and Stark effects
%32.10.-f Properties of atoms
%42.50.Lc Quantum fluctuations, quantum noise, and quantum jumps
% 33.55.Ad   Optical activity, optical rotation; circular dichroism
% 42.62.Fi   Laser Spectroscopy
% 33.55.Fi   Other magneto-optical and electro-optical effects
%07.55.Ge Magnetometers for magnetic field measurements

\maketitle

Recently there has been considerable interest in improving the
sensitivity of magnetic field measurements using techniques
associated with quantum entanglement and spin-squeezing.  In fact,
a recent experiment \cite{Ger2004} reported an atomic magnetometer
with noise below the shot-noise limit. This work utilized a
quantum nondemolition (QND) measurement of atomic spins---an
optical-rotation measurement with off-resonant light---to achieve
spin squeezing (see, for example, Refs.
\cite{Kuz2000,Smi2003,Ger2003} and references therein). The
purpose of this technique is to reduce the influence of the
quantum-mechanical spin-projection noise.

Here we analyze the sensitivity of an idealized magnetometer based
on these ideas and determine the scaling of the sensitivity with
various key parameters of the system (e.g., the number of atoms
$N$ and the measurement time $\tau$).  For concreteness, we
consider a magnetometer scheme (Fig.~1) in which a circularly
polarized pump beam orients $N$ paramagnetic atoms along $\uv{x}$.
%--------------------------------------------------------------------------
\begin{figure}
    \includegraphics[width=3 in]{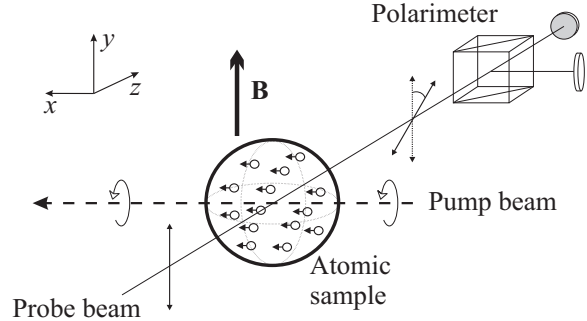}
    \caption{Schematic diagram of a typical atomic magnetometer
    apparatus \cite{Ger2004,Kom2003} of the sort considered
    here.}\label{Fig1}
\end{figure}
%--------------------------------------------------------------------------
When the pump beam is turned off, the atomic spins precess around
the direction of the magnetic field to be measured, assumed here
to be along $\uv{y}$. The spin precession is detected using
optical rotation of a far-detuned ($\abs{\Delta} \gg \Gamma_0$,
where $\Gamma_0$ is the natural transition width and $\Delta$ is
the frequency detuning from optical resonance), linearly polarized
probe beam propagating along $\uv{z}$, with cross-section (of area
$A$) assumed to match that of the atomic sample.

According to general principles of quantum mechanics, a
measurement perturbs the quantum state of the system under
observation. It is important to realize, however, that if one is
not attempting to extract the complete information about the
system, it is quite straightforward to set up a QND measurement
that will not alter the quantity one is trying to determine (see,
for example, Ref.\ \cite{Scu97}). Specifically, in the case
considered here, the orientation of the atomic spins in a given
direction is measured via optical rotation of the probe beam. If a
photon is absorbed from the probe light beam, the atom is excited
from the state one is attempting to measure and the orientation of
the system is altered. The photon-absorption probability scales
with detuning as $1/\Delta^2$, while optical rotation due to the
imbalance of the number of atoms oriented along and opposite to
the light-propagation direction scales as $1/\Delta$. Thus, an
approximation to a QND measurement of orientation is realized by
simply tuning the light sufficiently far away from resonance.
However, the residual absorption turns out to be important in
optimizing the measurement, as discussed below.

We assume that the pump beam prepares the $N$ paramagnetic atoms
with all spins polarized in the $\uv{x}$-direction. Without loss
of generality, we can assume that the magnetic field to be
detected is arbitrarily small. The measurement of the optical
rotation is carried out over a short time $\tau$ (the meaning of
``short'' will be specified more precisely below). In order to
make our argument as transparent as possible and to simplify the
mathematical expressions, in the following we neglect numerical
constants of order unity and set $\hbar=c=1$.

First, we recall the principle of the magnetometer's operation.
The polarized atoms undergo Larmor precession in the magnetic
field \footnote{If the ground-state angular momentum $J$ is
greater than $1/2$, more complicated polarization evolution will
also occur (in addition to the Larmor precession) due to the
light-induced Stark shifts, possibly contributing to the noise.
For simplicity, we will assume $J=1/2$ here.}, tipping their
polarization direction from the initial $\uv{x}$-direction towards
$\uv{z}$ by an angle $g\mu B\tau$ during the measurement time
$\tau$. Here $g$ is the Land\'{e} factor and $\mu$ is the Bohr
magneton. The angle of optical rotation induced by the excess of
atoms with spin projection along $\uv{z}$ can be written as
%--------------------------------------------------------------------------
\begin{equation}
    \varphi=g\mu B \tau \frac{l}{l_0} \frac{\Gamma_0}{\Delta}\,.
    \label{Eq:basic_Eq_for_phi}
\end{equation}
%--------------------------------------------------------------------------
Here $l$ is the length of the sample in the direction of the light
propagation, $l_0$ is the on-resonance unsaturated absorption
length, and the expression assumes far-detuned light and a weak
magnetic field.

There are two fundamental sources of noise that limit how well one
can determine $B$ from Eq.\ \eqref{Eq:basic_Eq_for_phi} (we assume
that the noise in the magnetic field is negligible). First, there
is photon shot noise in the optical polarimeter
%--------------------------------------------------------------------------
\begin{equation}
    \delta\varphi_{\rm ph}=\frac{1}{\sqrt{N_{\rm ph}}}\,,
    \label{Eq:photon_shot_noise}
\end{equation}
%--------------------------------------------------------------------------
where $N_{\rm ph}$ is the total number of photons used in the
measurement. From Eq.\ \eqref{Eq:basic_Eq_for_phi}, this noise
translates into the magnetic-field detection limit
%--------------------------------------------------------------------------
\begin{equation}
    \delta B_{\rm ph}
    =\frac{1}{g\mu\tau}
        \frac{1}{N\sqrt{N_{\rm ph}}}
        \frac{\Delta}{\Gamma_0}
        \frac{A}{\lambda^2}\,,
    \label{Eq:B2_noise_phot}
\end{equation}
%--------------------------------------------------------------------------
where we have written the resonant absorption length $l_0$ (which
can be thought of as a mean free path for a resonant photon) as
%--------------------------------------------------------------------------
\begin{equation}
    l_0=\frac{1}{n\lambda^2}=\frac{Al}{N\lambda^2}\,, \label{Eq:l_0}
\end{equation}
%--------------------------------------------------------------------------
where $n$ is the number density of the atoms and $\lambda$ is the
light wavelength.

The other source of noise is related to the fact that, even though
the probe light is far detuned from resonance, it still excites a
number $N_{\rm e}$ of atoms, given by the product of the resonant
excitation rate, a scaling factor taking into account the large
light detuning, and $N\tau$:
%--------------------------------------------------------------------------
\begin{equation}
    N_{\rm e}
    =\frac{d^2E^2}{\Gamma_0}\prn{\frac{\Gamma_0}{\Delta}}^2N\tau
    =N N_{\rm ph}\prn{\frac{\Gamma_0}{\Delta}}^2\frac{\lambda^2}{A}\,,
    \label{Eq:N_exc}
\end{equation}
%--------------------------------------------------------------------------
where $d$ is the dipole moment of the probe transition and $E$ is
the amplitude of the probe-light field. Here we have used
$d^2=\Gamma_0\lambda^3$ and
%--------------------------------------------------------------------------
\begin{equation}
    E^2=\frac{\Phi}{\lambda}=\frac{N_{\rm ph}}{\lambda A
    \tau}\,,\label{Eq:E2}
\end{equation}
%--------------------------------------------------------------------------
where $\Phi=N_{\rm ph}/(A\tau)$ is the photon flux. Such
excitation results in a random imbalance of order $\sqrt{N_{\rm
e}}$ between the number of atoms with positive and negative
spin-projection along $\uv{z}$. This leads to optical rotation by
a random angle
\begin{equation}
    \delta\varphi_{\rm at}
    \sim\frac{\sqrt{N_{\rm
    e}}}{N}\frac{l}{l_0}\frac{\Gamma_0}{\Delta}\,,
    \label{Eq:rot due to pumped atoms}
\end{equation}
and a corresponding uncertainty
\begin{equation}
    \delta B_{\rm at}
    =\frac{1}{g\mu\tau}
        \frac{\sqrt{N_{\rm ph}}}{\sqrt{N}}
        \frac{\Gamma_0}{\Delta}
        \prn{\frac{\lambda^2}{A}}^{1/2}.
    \label{Eq:B2_noise_at}
\end{equation}
It is important to emphasize that the uncertainty in
magnetic-field determination described by Eq.\
\eqref{Eq:B2_noise_at} arises solely due to optical pumping
induced by the \emph{probe} beam during the measurement time
$\tau$. As shown in Refs.\ \cite{Ger2004,Ger2003}, the projection
noise due to the initial spin preparation can be eliminated by use
of a proper measurement procedure.

We see that the two contributions to uncertainty in the
magnetic-field determination---one associated with the polarimeter
photon noise, the other associated with reorientation of atoms by
the probe light---have opposite dependences on $N_{\rm ph}$. We
can find the optimum number of photons by minimizing the overall
uncertainty. Differentiating the sum in quadrature of the
contributions of Eqs.\ \eqref{Eq:B2_noise_phot} and
\eqref{Eq:B2_noise_at} by $N_{\rm ph}$ and setting the derivative
to zero, we find the optimal value
\begin{equation}
    N_{\rm ph}^{\rm opt}
        =\frac{1}{N^{1/2}}
        \prn{\frac{\Delta}{\Gamma_0}}^2
        \prn{\frac{A}{\lambda^2}}^{3/2},
    \label{Eq:N_phot_opt}
\end{equation}
for which the photon and atomic noise contributions are the same.
The resultant overall uncertainty in the determination of the
magnetic field in a single measurement of length $\tau$ is
\begin{equation}
    \delta B
    =\frac{1}{g\mu\tau}
        \frac{1}{N^{3/4}}
        \prn{\frac{A}{\lambda^2}}^{1/4}.
    \label{Eq:delta B optimized}
\end{equation}
Note that the transition line width and frequency detuning have
dropped out of the optimized result \eqref{Eq:delta B optimized}.
Equation \eqref{Eq:delta B optimized} shows that the sensitivity
to the magnetic field scales as $N^{-3/4}$, better than the
scaling $N^{-1/2}$ for a usual shot-noise-limited measurement
\cite{Bud2002RMP,Kom2003}, but still short of the result $N^{-1}$
obtained in the Heisenberg limit. The factor $(A/\lambda^2)^{1/4}$
indicates that, given a total number of atoms $N$, it is
beneficial to compress their dimensions down to the wavelength of
the light, maximizing the optical rotation angle. This, however,
may be difficult to achieve experimentally, and may also lead to
cooperative effects, not considered here, in the light-atom
interaction.

It is interesting to note that with an optimized measurement the
number of atoms that undergo optical pumping during the
measurement time $\tau$ is [using Eq.\ \eqref{Eq:N_phot_opt}]
%--------------------------------------------------------------------------
\begin{equation}
\frac{d^2E^2}{\Delta^2}\Gamma_0\tau N=\lambda^{-1}
A^{1/2}\sqrt{N}. \label{Eq:N pumped}
\end{equation}
%--------------------------------------------------------------------------

Next, we consider the possibility of improving magnetometric
sensitivity by employing strongly squeezed probe light
\cite{Scu97}. In this case, the photon noise contribution
approaches $1/N_{\rm ph}$ [cf. Eq.\ \eqref{Eq:photon_shot_noise}]
when 100\%-efficient photodetection is assumed; the minimization
of the uncertainty in the magnetic-field determination leads to
the optimal number of photons
\begin{equation}
    N_{\rm ph}^{\rm opt}
    =\frac{1}{N^{1/3}}
        \prn{\frac{\Delta}{\Gamma_0}}^{4/3}
        \frac{A}{\lambda^2}\,,
    \label{Eq:Phi optimized_squeezed}
\end{equation}
and uncertainty in magnetic-field detection of
\begin{equation}
    \delta B
    =\frac{1}{g\mu\tau}
        \frac{1}{N^{2/3}}
        \prn{\frac{\Gamma_0}{\Delta}}^{1/3}.
    \label{Eq:delta B optimized_squeezed}
\end{equation}
In contrast to the case of unsqueezed light [Eq.\ \eqref{Eq:delta
B optimized}], the detuning $\Delta$ has not cancelled, while the
area $A$ has. This seems to be an improvement on both fronts. To
obtain the greatest sensitivity, the atomic sample no longer needs
to be compressed to the scale of the light wavelength. Also, it
would appear that $\delta B$ can be decreased without limit by
increasing the detuning. However, there is a fundamental limit to
the sensitivity that can be derived from the Heisenberg
uncertainty principle:
%--------------------------------------------------------------------------
\begin{equation}
    \delta B_H
    =\frac{1}{g\mu\tau}
    \frac{1}{N}\,.
    \label{Eq:delta B H}
\end{equation}
%--------------------------------------------------------------------------
Equating \eqref{Eq:delta B optimized_squeezed} and \eqref{Eq:delta
B H}, we find that the Heisenberg limit is achieved when
$\Delta=N\Gamma_0$. Putting this value of the detuning into Eqs.\
\eqref{Eq:Phi optimized_squeezed} and \eqref{Eq:N_exc}, we see
that the optimal number of photons is $N_{\rm ph}^{\rm
opt}=NA/\lambda^2$ and the number of atoms optically pumped during
the measurement is of order unity (i.e., their relative fraction
is $1/N$, independent of any parameters of the optimized system).
Indeed, since the Heisenberg limit is reached when the change of
one atomic spin due to the magnetic field can be measured, a
greater number of spins must not be disturbed by the light.

In order to obtain the greatest sensitivity to magnetic fields, it
is advantageous to make a single QND measurement over as long a
time as possible. Up until now, we have ignored the ground-state
spin relaxation (with rate $\Gamma_{\rm rel}$), as we have assumed
$\tau$ sufficiently short. For longer measurement times, the
approximation that the spins reorient only due to optical pumping
by the probe beam will fail. For the case of a measurement using
squeezed light, when the number of spins ($N\Gamma_{\rm rel}\tau$)
that flip due to ground-state relaxation becomes comparable to
unity, uncertainty due to relaxation begins to dominate the atomic
noise. The additional noise in the optical rotation angle due to
the relaxation of $N\Gamma_{\rm rel}\tau$ atoms during the
measurement is given, analogously to Eq.\ \eqref{Eq:rot due to
pumped atoms}, by
%--------------------------------------------------------------------------
\begin{equation}
    \delta\varphi_{\rm rel}
    \sim\frac{\sqrt{N\Gamma_{\rm rel}\tau}}{N}
        \frac{l}{l_0}
        \frac{\Gamma_0}{\Delta}\,.
\end{equation}
%--------------------------------------------------------------------------
The corresponding noise in the magnetic-field determination is
given by
%--------------------------------------------------------------------------
\begin{equation}
    \delta B_{\rm rel}
    =\frac{\sqrt{\Gamma_{\rm rel}}}{g\mu\sqrt{N\tau}}\,.
    \label{Eq:delta B relaxation}
\end{equation}
%--------------------------------------------------------------------------
Thus it is evident that if the measurement is performed over a
time $\tau \gg (N\Gamma_{\rm rel})^{-1}$ any advantage in
sensitivity due to the QND measurement is lost since the noise
scales as $N^{-1/2}$. For a total measurement time $T \gg
(N\Gamma_{\rm rel})^{-1}$, one can perform  $T/\tau = N\Gamma_{\rm
rel}T$ independent Heisenberg-limited measurements of the magnetic
field, each of duration $\tau = (N\Gamma_{\rm rel})^{-1}$ and
sensitivity $\Gamma_{\rm rel}/(g\mu)$ [Eq.\ \eqref{Eq:delta B H}].
The total uncertainty in the measurement improves as the square
root of the number of such independent measurements.  Thus the
sensitivity achieved during the measurement time is given by
%--------------------------------------------------------------------------
\begin{equation}
    \delta B = \frac{\sqrt{\Gamma_{\rm rel}}}{g\mu\sqrt{NT}}\,,
    \label{Eq:delta B long T}
\end{equation}
%--------------------------------------------------------------------------
which is the same as the sensitivity of a conventional
shot-noise-limited magnetometer.

A similar conclusion is also reached for the case of unsqueezed
probe light. Here, the maximal measurement time during which no
relaxation events that would spoil the sensitivity can occur is
%--------------------------------------------------------------------------
\begin{equation}
    \tau=\frac{1}{\sqrt{N}}\frac{1}{\Gamma_{\rm rel}}\prn{\frac{A}{\lambda^2}}^{1/2},
\end{equation}
%--------------------------------------------------------------------------
which once again leads us to the result \eqref{Eq:delta B long T}.
A similar result was obtained in the context of frequency
measurements in the presence of decoherence \cite{Hue97}, where it
was shown that optimal measurements with maximally entangled
states offer no improvement over standard spectroscopic
techniques.

The preceding analysis suggests the general result that, while it
is possible to perform measurements that go beyond the shot-noise
limits for very short times, the inevitable presence of
ground-state relaxation means that the most sensitive
measurements---requiring longer measurement times---will have the
usual shot-noise scaling of the sensitivity. As a numerical
example, Heisenberg-limited measurements for $N=10^{11}$ and
$\Gamma_{\rm rel}=100$ Hz (parameters comparable to those used in
Ref.\ \cite{Ger2004}) must be shorter than $10^{-13}$ s.

So far, we have considered two fundamental limits to the
magnetometric sensitivity: photon shot noise and optical pumping
by the probe light. In addition to these sources of noise, the
probe beam also contributes noise due to quantum fluctuations of
its polarization.  This leads to a differential AC Stark shift of
the ground state magnetic sublevels. Although this effect does not
change the scaling of the magnetometric precision, as we show
below, it is important to account for such noise when considering
the Heisenberg uncertainty relations for the atomic spins.

Although the probe beam is nominally linearly polarized, vacuum
fluctuations in the orthogonal polarization can create a small
admixture of random circular polarization.  The magnitude of the
quantum fluctuations of the probe polarization can be found using
the ellipticity operator $\hat{\epsilon}$ for nominally
$y$-polarized light \cite{Mat2002}
%--------------------------------------------------------------------------
\begin{align}
\hat{\epsilon} =
\frac{{\mathcal{E}_0}}{2iE}\prn{\hat{a}_x-\hat{a}_x^\dag}~,
\end{align}
%--------------------------------------------------------------------------
where ${\mathcal{E}}_0$ is the characteristic amplitude of
unsqueezed vacuum fluctuations (see, for example, Ref.
\cite{Scu97}) and $\hat{a}_x$, $\hat{a}_x^\dag$ are the
annihilation and creation operators for $x$-polarized photons at
the same frequency as the probe. Assuming that the $x$-polarized
field is the unsqueezed vacuum, we find for the quantum
fluctuations of the probe beam's ellipticity (details of the
calculation are presented in Ref. \cite{Auz2004App}):
%--------------------------------------------------------------------------
\begin{align}
\delta\epsilon =
\sqrt{\abrk{\hat{\epsilon}^2}-\abrk{\hat{\epsilon}}^2} =
\frac{1}{\sqrt{N_{\rm ph}}}~. \label{Eq:epsilon}
\end{align}
%--------------------------------------------------------------------------
The magnitude of the differential AC Stark shift of the ground
state magnetic sublevels, $\delta\Delta_{ac}$, due to the
fluctuations of the probe polarization is
%--------------------------------------------------------------------------
\begin{equation}
\delta\Delta_{ac} =
\frac{d^2E^2}{\Delta}\delta\epsilon.\label{Eq:AC Stark}
\end{equation}
%--------------------------------------------------------------------------
This causes the atomic polarization vector to precess by a random
angle in the x-y plane (Fig. \ref{Fig1}). This small-angle
precession in the x-y plane does not affect the magnetometric
sensitivity. After time $\tau$ this random angle has a magnitude
$\alpha_{x-y} = \tau\delta\Delta_{ac}$. After substitutions from
Eqs.\ \eqref{Eq:E2}, \eqref{Eq:N_phot_opt}, \eqref{Eq:epsilon},
and \eqref{Eq:AC Stark}, the following expression is obtained:
%--------------------------------------------------------------------------
\begin{equation}
\alpha_{x-y} =
\frac{1}{N^{1/4}}\left(\frac{A}{\lambda^2}\right)^{-1/4}.\label{Eq:alphaXY}
\end{equation}
%--------------------------------------------------------------------------

This rotation of the atomic polarization vector in the x-y plane
ensures that the measurement uncertainties obey the Heisenberg
uncertainty relation
%--------------------------------------------------------------------------
\begin{equation}
\delta J_z\delta J_y \geq J_x.\label{Eq:Heisenberg}
\end{equation}
%--------------------------------------------------------------------------
Let us verify this. From Eq.\ \eqref{Eq:delta B optimized} we
obtain
\begin{align}
\delta J_z = N\tau g\mu\delta B = N^{1/4} (A/\lambda^2)^{1/4}~.
\end{align}
Using Eq.\ \eqref{Eq:alphaXY},
\begin{align}
\delta J_y = N\alpha_{x-y} = N^{3/4}(A/\lambda^2)^{-1/4}~.
\end{align}
To first order, $J_x = N$, so the uncertainty relation
\eqref{Eq:Heisenberg} is satisfied.

In the case of squeezed probe light, the random admixture of
circular polarization due to quantum fluctuations is in fact
$\delta\epsilon \sim 1$; this result can be derived in the same
manner as Eq.\ \eqref{Eq:epsilon}, except here one uses a squeezed
vacuum state (ensuring the photon noise in the probe optical
rotation measurement at the $1/N_{\rm ph}$ level) for the
$x$-polarized field. Repeating the AC Stark effect calculation
(with the factor $A/\lambda^2$ set to unity) gives the angle of
rotation in the x-y plane $\alpha_{x-y} \sim 1$, which means that
$\delta J_y = J_x = N$. But at the Heisenberg limit $\delta J_z =
1$, so once again the uncertainty relation \eqref{Eq:Heisenberg}
holds.

In conclusion, we have investigated fundamental sources of noise
present in an idealized atomic magnetometer based on quantum
non-demolition techniques.  We find that such an approach can
improve the sensitivity of magnetometric measurements beyond the
shot-noise-limit over time scales much shorter than the relevant
spin relaxation time divided by an appropriate power of the number
of atoms, depending on the degree of squeezing of the probe light.
However, for longer time scales, even if squeezed probe light is
employed, QND techniques offer no significant improvement in the
sensitivity of magnetic field measurements.

The authors are grateful to J. M. Geremia, M. Graf, J. Guzman, E.
Williams, M. Romalis, and M. Zolotorev for useful discussions.
This work was supported by NSF, a NATO linkage grant, and by the
Director, Office of Science, Office of Basic Energy Sciences,
Materials Sciences and Nuclear Science Divisions, of the U.S.
Department of Energy under contract DE-AC03-76SF00098.

\bibliography{NMObibl}

\end{document}